\documentstyle[12pt,amssymb,amstex]{amsart}

\numberwithin{equation}{section}

\newcommand{\bea}{\begin{eqnarray}}
\newcommand{\eea}{\end{eqnarray}}
\newcommand{\ba}{\begin{array}}
\newcommand{\ea}{\end{array}}
\newcommand{\edc}{\end{document}}
\newcommand{\bc}{\begin{center}}
\newcommand{\ec}{\end{center}}
\newcommand{\be}{\begin{equation}}
\newcommand{\ee}{\end{equation}}

\newcommand{\dsf}{\displaystyle\frac}

\def\cb{{\cal B}}

\def\cf{{\cal F}}
\def\cg{{\cal G}}

\def\cs{{\cal S}}

\def\cw{{\cal W}}

\def\gh{{\frak h}}

\def\bc{{\mathbb C}}

\def\bn{{\mathbb N}}

\def\bq{{\mathbb Q}}

\def\bz{{\mathbb Z}}

  \def\G{\Gamma}
\def\d{\delta}  

\def\e{\epsilon}
\def\l{\lambda} 
\def\k{\kappa}
\def\m{\mu}

\def\s{\sigma} 
\def\t{\theta}

\def\w{\omega} \def\Om{\Omega}

\def\h{{\mathbf{h}}}
\def\sb{{\mathbf{s}}}
\def\xb{{\mathbf{x}}}
\def\yb{{\mathbf{y}}}
\def\Bb{{\mathbf{B}}}

\newtheorem{thm}{Theorem}[section]
\newtheorem{lem}[thm]{Lemma}

\theoremstyle{remark}
\newtheorem{rem}{Remark}[section]
\newtheorem{ex}{Example}[section]

\begin{document}
\small

\title[$p$-adic Gibbs measure of Potts model]
{On $p$-adic Gibbs measures of countable state Potts model on the
Cayley tree}


\author{A.Yu. Khrennikov}
\address{A. Yu. Khrennikov\\
International Center for Mathematical Modeling\\
MSI, V\"{a}xj\"{o} University\\ SE-35195,  V\"{a}xj\"{o},
Sweden}\email{{\tt Andrei.Khrennikov@@msi.vxu.se}}
\author{F. M. Mukhamedov}
\address{F.M. Mukhamedov\\
Departamento de Fisica,
Universidade de Aveiro\\
Campus Universitario de Santiago\\
3810-193 Aveiro, Portugal} \email{{\tt far75m@@yandex.ru}, {\tt
farruh@@fis.ua.pt}}
\author{J. F.F. Mendes}
\address{J. F.F. Mendes\\
Departamento de Fisica \\
Universidade de Aveiro\\
Campus Universit\'{a}rio de Santiago\\
3810-193 Aveiro, Portugal} \email{{\tt jfmendes@@fis.ua.pt}}

\begin{abstract}
In this paper we consider countable state $p$-adic Potts model on
the Cayley tree. A construction of $p$-adic Gibbs measures which
depends on weights $\l$ is given, and an investigation of such
measures is reduced to examination of an infinite-dimensional
recursion equation. Studying the derived equation under some
condition concerning weights, we prove the absence of a phase
transition. Note that the condition does not depend on values of
the prime $p$, and an analogues fact is not true when the number
of spins is finite. For homogeneous model it is shown that the
recursive equation has only one solution  under that condition on
weights. This means that there is only one $p$-adic Gibbs measure
$\m_\l$. The boundedness of the measure is also established.
Moreover, continuous dependence of the measure $\m_\l$ on $\l$ is
proved. At the end we formulate one limit theorem for $\m_\l$.

\end{abstract}

\maketitle

\footnotetext[1]{ Current address (F.M.): Department of Comput. \&
Theor. Sci., Faculty of Sciences, IIUM, P.O. Box, 141, 25710,
Kuantan, Pahang, Malaysia }
\section{introduction}

Since  the 1980s, various models described in the language of
$p$-adic analysis have been actively studied
\cite{ADFV},\cite{FO,FW},\cite{MP},\cite{V1}. More precisely,
models defined over the field of $p$-adic numbers have been
considered. Which is due to the assumption that $p$-adic numbers
provide a more exact and more adequate description of microworld
phenomena. The well-known studies in this area are primarily
devoted to investigating quantum mechanics models using equations
of mathematical physics \cite{ADV,M,PM,V2,VVZ}. One of the first
applications of $p$-adic numbers in quantum physics appeared in
the framework of quantum logic in \cite{BC}. This model is
especially interesting for us because it could not be described by
using conventional real valued probability. Furthermore, numerous
applications of the $p$-adic analysis to mathematical physics have
been proposed in \cite{ABK},\cite{Kh1},\cite{Kh2},\cite{PM}.
Besides, it is also known \cite{Kh2,Ko,MP,PM,Ro,Vi,VVZ} that a
number of $p$-adic models in physics cannot be described using
ordinary Kolmogorov's probability theory. New probability models,
namely $p$-adic ones were investigated in
\cite{BD},\cite{Br},\cite{K3},\cite{KhN}. After that in \cite{KYR}
an abstract $p$-adic probability theory was developed by means of
the theory of non-Archimedean measures \cite{Ro}. Using that
measure theory in \cite{KL},\cite{Lu} the theory of stochastic
processes with values in $p$-adic and more general non-Archimedean
fields having probability distributions with non-Archimedean
values has been developed. In particular, a non-Archimedean analog
of the Kolmogorov theorem was proven (see also \cite{GMR}). Such a
result allows us to construct wide classes of stochastic processes
using finite dimensional probability distributions. We point out
that stochastic processes on the field $\bq_p$ of $p$-adic numbers
have been studied by many authors, for example,
\cite{AK,AZ1,AZ2,DF,Koc,Y}. In those investigations  wide classes
of Markov processes on $\bq_p$ were constructed and studied. Such
studies, therefore, give a possibility to develop the theory of
statistical mechanics in the context of the $p$-adic theory, since
it lies on the basis of the theory of probability and stochastic
processes. Note that one of the central problems of such a theory
is the study of infinite-volume Gibbs measures corresponding to a
given Hamiltonian, and a description of the set of such measures.
In most cases such an analysis depend on a specific properties of
Hamiltonian, and complete description is often a difficult
problem. This problem, in particular, relates to a phase
transition of the model (see \cite{G}).

In \cite{KK1,KK2} a notion of ultrametric Markovianity, which
describes independence of contributions to random field from
different ultrametric balls, has been introduced, and shows that
Gaussian random fields on general ultrametric spaces (which were
related with hierarchical trees), which were defined as a solution
of pseudodifferential stochastic equation (see also \cite{KaKo}),
satisfies the Markovianity. In addition,  covariation of the
defined random field was computed with the help of wavelet
analysis on ultrametric spaces (see also \cite{Koz}). Some
applications of the results to replica matrices, related to
general ultrametric spaces have been investigated in \cite{KK3}.

In this paper we develop a $p$-adic probability theory approach to
study countable state of nearest-neighbor Potts models on a Cayley
tree (see \cite{B}) over $p$-adic filed. We are especially
interested in the construction of $p$-adic Gibbs measure for the
mentioned model. Such measures present more natural concrete
examples of $p$-adic Markov processes (see \cite{KK2,KL}, for
definitions). When states are finite, say $q$, then the
corresponding $p$-adic $q$-state Potts models on the same tree
have been studied in \cite{MR1,MR2,MRM}\footnote{The classical
(real value) counterparts of such models were considered in
\cite{GR}, \cite{W}}. It was established that a phase transition
occurs if $q$ is divisible by $p$. This shows that the transition
depends on the number of spins $q$. To establish such a result we
investigated fixed points of a $p$-adic dynamical systems
associated with that model. We remark that first investigations of
non-Archimedean dynamical systems have appeared in \cite{HY}. We
also point out that intensive development of $p$-adic (and more
general algebraic) dynamical systems has happened few years, see
\cite{AV1,AV2,B1,Be,KhN,Sil2,TVW,Wa}. More extensive lists may be
found in the $p$-adic dynamics bibliography maintained by
Silverman \cite{Sil2} and the algebraic dynamics bibliography of
Vivaldi \cite{Vi}.

The aim of this paper is to give sufficient condition for the
uniqueness of $p$-adic Gibbs measures of the countable state Potts
model, and to study such measures. Note that in comparison to a
real case, in a $p$-adic setting, \`{a} priori the existence of
such kind of measures for the model is not known, since there is
not much information on topological properties of the set of all
$p$-adic measures defined even on compact spaces. However, in the
real case, there is the so called the Dobrushin's Theorem
\cite{Dob1,Dob2} which gives a sufficient condition for the
existence of the Gibbs measure for a large class of Hamiltonians.

The paper is organized as follows. After preliminaries (Sec. 2) in
section 3 we define our model, and give a construction of $p$-adic
Gibbs measures which depends on weight $\l$. Using Kolmogorov
extension Theorem \cite{KL}, an investigation of such measures is
reduced to examination of an infinite-dimensional recursion
equation. In Section 4, studying the derived equation under some
condition on weights, we prove the absence of a phase transition.
Note that, for the real counterparts of the model, such results
are unknown (see \cite{Ga,GR}). It turns out that the founding
condition does not depend on values of the prime $p$, and
therefore, an analogous fact is not true when the number of spins
is finite. In section 5, we well consider homogeneous $p$-adic
Potts model, and show under the condition formulated in section 3,
the recursive equation has only one solution. Hence, there is only
one $p$-adic Gibbs measure $\m_\l$. Then we establish boundedness
one. In section 6, we prove continuous dependence of the measure
$\m_\l$ on $\l$. We also prove one limit theorem for $\m_\l$. The
last section is devoted to the conclusions of the results.

\section{Preliminaries}

In what follows $p$ will be a fixed prime number, and $\bq_p$
denotes the field of $p$-adic filed, formed by completing $\bq$
with respect to the unique absolute value satisfying $|p|_p =
1/p$. The absolute value $|\cdot|_p$, is non- Archimedean, meaning
that it satisfies the ultrametric triangle inequality $|x + y|_p
\leq \max\{|x|_p, |y|_p\}$.

Given $a\in \bq_p$ and  $r>0$ put
$$
B(a,r)=\{x\in \bq_p : |x-a|_p< r\}.
$$
The {\it $p$-adic logarithm} is defined by the series
$$
\log_p(x)=\log_p(1+(x-1))=\sum_{n=1}^{\infty}(-1)^{n+1}\dsf{(x-1)^n}{n},
$$
which converges for $x\in B(1,1)$; the {\it $p$-adic exponential}
is defined by
$$
\exp_p(x)=\sum_{n=0}^{\infty}\dsf{x^n}{n!},
$$
which converges for $x\in B(0,p^{-1/(p-1)})$.

\begin{lem}\label{exp}\cite{Ko,S} Let $x\in B(0,p^{-1/(p-1)})$
then we have
\begin{eqnarray}\label{exp}
&& |\exp_p(x)|_p=1,\ \ \ |\exp_p(x)-1|_p=|x|_p,
\ \ \ |\log_p(1+x)|_p=|x|_p \\
\label{el} && \log_p(\exp_p(x))=x, \ \ \exp_p(\log_p(1+x))=1+x.
\end{eqnarray}
\end{lem}

Note the basics of $p$-adic analysis, $p$-adic mathematical
physics are explained in \cite{Ko,S,VVZ}.

Let $(X,{\cal B})$ be a measurable space, where ${\cal B}$ is an
algebra of subsets $X$. A function $\m:{\cal B}\to \bq_p$ is said
to be a {\it $p$-adic measure} if for any
$A_1,\dots,A_n\subset{\cal B}$ such that $A_i\cap A_j=\emptyset$
($i\neq j$) the equality holds
$$
\mu\bigg(\bigcup_{j=1}^{n} A_j\bigg)=\sum_{j=1}^{n}\mu(A_j).
$$

A $p$-adic measure is called a {\it probability measure} if
$\mu(X)=1$. A $p$-adic probability measure $\m$ is called {\it
bounded} if $\sup\{|\m(A)|_p : A\in \cb\}<\infty $. Note that in
general, a $p$-adic probability measure need not be bounded
\cite{K3,KL,Ko}. For more detail information about $p$-adic
measures we refer to \cite{K3},\cite{KhN},\cite{Ro}.

Recall that the Cayley tree (see \cite{B}) $\Gamma^k$ of order $
k\geq 1 $ is an infinite tree, i.e., a graph without cycles, from
each vertex of which exactly $ k+1 $ edges issue. Let
$\Gamma^k=(V, L)$, where $V$ is the set of vertexes of $
\Gamma^k$, $L$ is the set of edges of $ \Gamma^k$. The vertices
$x$ and $y$ are called {\it nearest neighbors} and they are
denoted by $l=<x,y>$ if there exists an edge connecting them. A
collection of the pairs $<x,x_1>,\dots,<x_{d-1},y>$ is called a
{\it path} from the point $x$ to the point $y$. The distance
$d(x,y), x,y\in V$, on the Cayley tree, is the length of the
shortest path from $x$ to $y$. Now fix $ x^0 \in V $, and set
$$ W_n=\{x\in V| d(x,x^0)=n\}, \ \ \
V_n=\bigcup_{m=1}^n W_m, \ \ L_n=\{l=<x,y>\in L | x,y\in V_n\}.
$$ The set of {\it direct successors} of $x$ is defined by
\begin{equation}\label{S(x)}
S(x)=\{y\in W_{n+1} :  d(x,y)=1 \}, \ \ x\in W_n.
\end{equation}
Observe that any vertex $x\neq x^0$ has $k$ direct successors and
$x^0$ has $k+1$.

\section{The $p$-adic Potts model and $p$-adic Gibbs measures}

We consider the $p$-adic Potts model where spin takes values in
the set $\Phi=\{0,1,2,\cdots\}$ ($\Phi$ is called a {\it state
space}) and is assigned to the vertices of the tree
$\G^k=(V,\Lambda)$. A configuration $\s$ on $V$ is then defined as
a function $x\in V\to\s(x)\in\Phi$; in a similar manner one
defines configurations $\s_n$ and $\w$ on $V_n$ and $W_n$,
respectively. The set of all configurations on $V$ (resp. $V_n$,
$W_n$) coincides with $\Omega=\Phi^{V}$ (resp.
$\Omega_{V_n}=\Phi^{V_n},\ \ \Omega_{W_n}=\Phi^{W_n}$). One can
see that $\Om_{V_n}=\Om_{V_{n-1}}\times\Om_{W_n}$. Using this, for
given configurations $\s_{n-1}\in\Om_{V_{n-1}}$ and
$\w\in\Om_{W_{n}}$ we define their concatenations  by
$$
(\s_{n-1}\vee\w)(x)= \left\{
\begin{array}{ll}
\s_{n-1}(x), \ \ \textrm{if} \ \  x\in V_{n-1},\\
\w(x), \ \ \ \ \ \ \textrm{if} \ \ x\in W_n.\\
\end{array}
\right.
$$
It is clear that $\s_{n-1}\vee\w\in \Om_{V_n}$.

The Hamiltonian $H_n:\Om_{V_n}\to\bq_p$ of the inhomogeneous
$p$-adic countable state Potts model has a form
\begin{equation}\label{Potts}
H_n(\s)=\sum_{<x,y>\in L_n}J_{x,y}\delta_{\s(x),\s(y)},  \ \
\s\in\Om_{V_n}, \  n\in\mathbb{N},
\end{equation}
where $\delta$ is the Kronecker symbol and the coupling constants
$J_{x,y}$ are taken from $\bq_p$ with constraint
\begin{equation}\label{J}
|J_{x,y}|_p<\frac{1}{p^{1/(p-1)}}, \ \ \ \forall <x,y>\in L_n.
\end{equation}
Note that such a condition provides the existence of a $p$-adic
Gibbs measure (see \eqref{mu}).

We say \eqref{Potts} is {\it homogeneous Potts model} if
$J_{xy}=J, \ \ \forall <x,y>.$

We then construct $p$-adic Gibbs measures corresponding to the
model.

A given set $A$ we put $\bq_p^{A}=\{\{x_i\}_{i\in A}:
x_i\in\bq_p\}.$

Assume that a function $\h: V\setminus\{x^{(0)}\}\to\bq_p^{\Phi}$,
i.e. $\h_x=\{h_{i,x}\}_{i\in\Phi}$, is such that
\begin{equation}\label{h_i}
|h_{i,x}|_p<\frac{1}{p^{1/(p-1)}} \ \ \textrm{for all}\ \ x\in
V\setminus\{x^{(0)}\},\ i\in\Phi,
\end{equation}
and  a non-zero element $\l=\{\l(i)\}_{i\in\Phi}\in\bq_p^{\Phi}$
is fixed such that
\begin{equation}\label{L}
|\l(n)|_p\to 0 \ \ \textrm{as} \ \ n\to\infty
\end{equation}
which is called a {\it weight}. In what follows, without losing
generality we may assume that $\l(0)\neq 0$.

Given $n=1,2,\dots$ a $p$-adic probability measure $\m^{(n)}_\h$
on $\Om_{V_n}$ is defined by
\begin{equation}\label{mu}
\mu^{(n)}_{\h}(\s)=\frac{1}{Z_n^{(\h)}}\exp_p\bigg\{H_n(\s)+\sum_{x\in
W_n}h_{\s(x),x}\bigg\}\prod_{x\in V_n}\l(\s(x))
\end{equation}
Here, $\s\in\Om_{V_n}$, and $Z_n^{(\h)}$ is the corresponding
normalizing factor called a {\it partition function} given by
\begin{equation}\label{ZN1}
Z_n^{(\h)}=\sum_{\s\in\Omega_{V_n}}\exp_p\bigg\{H_n(\s)+\sum_{x\in
W_n}h_{\s(x),x}\bigg\}\prod_{x\in V_n}\l(\s(x)),
\end{equation}
here subscript $n$ and superscript $(\h)$ are accorded to the $Z$,
since it depends on $n$ and a function $\h$.

 The conditions \eqref{J} and \eqref{h_i} allow the existence
of $\exp_p$,  therefore, the measures $\m^{(n)}_\gh$ are well
defined.

Now we want to define a $p$-adic probability measure $\m$ on $\Om$
such that it would be compatible with defined ones $\m_\h^{(n)}$,
i.e.
\begin{equation}\label{CM}
\m(\s\in\Om: \s|_{V_n}=\s_n)=\m^{(n)}_\h(\s_n), \ \ \ \textrm{for
all} \ \ \s_n\in\Om_{V_n}, \ n\in\bn.
\end{equation}

In general, \`{a} priori the existence of such kind of measure
$\m$ is not known, since, there is not much information on
topological properties, such as compactness, of the set of all
$p$-adic measures defined even on compact spaces\footnote{In the
real case, when the state space is compact, then the existence
follows from the compactness of the set of all probability
measures (i.e. Prohorov's Theorem). When the state space is
non-compact, then there is a Dobrushin's Theorem \cite{Dob1,Dob2}
which gives a sufficient condition for the existence of the Gibbs
measure for a large class of Hamiltonians. In \cite{Ga} using that
theorem it has been established the existence of the Gibbs measure
for the real counterpart of the studied Potts model. It should be
noted that there are even nearest-neighbor models with countable
state space for which the Gibbs measure does not exists
\cite{Sp}.}. Therefore, at a moment, we can only use the so called
{\it compatibility condition} for the measures $\m_\h^{(n)}$,
$n\geq 1$, i.e.
\begin{equation}\label{comp}
\sum_{\w\in\Om_{W_n}}\m^{(n)}_\h(\s_{n-1}\vee\w)=\m^{(n-1)}_\h(\s_{n-1}),
\end{equation}
for any $\s_{n-1}\in\Om_{V_{n-1}}$ (cp. \cite{BRZ}), which implies
the existence of a unique $p$-adic measure $\m$ defined on $\Om$
with a required condition \eqref{CM}. Moreover, if the measures
$\m_\h^{(n)}$ are bounded, then $\m$ is also bounded. This
assertion is known as the $p$-adic Kolmogorov extension Theorem
(see \cite{GMR},\cite{KL}).

So, if for some function $\h$ the measures  $\m_\h^{(n)}$ satisfy
the compatibility condition, then there is a unique $p$-adic
probability measure, which we denote by $\m_\h$, since it depends
on $\h$. Such a measure $\m_\h$ is said to be {\it $p$-adic Gibbs
measure} corresponding to the $p$-adic Potts model. By ${\cal S}$
we denote the set of all such $p$-adic Gibbs measures. If there
are at least two different $p$-adic Gibbs measures in ${\cal S}$,
i.e. one can find two different functions $\sb$ and $\h$ defined
on $V\setminus\{x^0\}$ such that there exist the corresponding
measures $\m_\sb$ and $\m_\h$, which are different, then we say
that {\it a phase transition} occurs for the model, otherwise,
there is {\it no phase transition}.

Now one can ask for what kind of functions $\h$ the measures
$\m_\h^{(n)}$ defined by \eqref{mu} would satisfy the
compatibility condition \eqref{comp}. The following theorem gives
an answer to this question.

\begin{thm}\label{comp1} The measures $\m^{(n)}_\h$, $
n=1,2,\dots$ given by \eqref{mu}, satisfy the compatibility
condition \eqref{comp} if and only if for any $x\in
V\setminus\{x^{(0)}\}$ the following equation holds:
\begin{equation}\label{eq1}
\hat h_{i,x}=\frac{\l(i)}{\l(0)}\prod_{y\in S(x)}F_i(\hat
\h_y;\theta_{x,y}), \ \ i\in\bn
\end{equation}
here and below $\theta_{x,y}=\exp_p(J_{x,y})$, a sequence $\hat
\h=\{\hat h_i\}_{i\in\bn}\in\bq_p^\bn$ is defined by 
$\h=\{h_i\}_{i\in\Phi}$ as follows
\begin{equation}\label{H}
\hat h_i=\exp_p(h_i-h_0)\frac{\l(i)}{\l(0)}, \ \ \ i\in\bn
\end{equation}
and mappings $F_i:\bq_p^{\bn}\times\bq_p\to\bq_p$ are defined by
\begin{equation}\label{eq2}
F_i(\xb;\theta)=\frac{(\theta-1)x_i+\sum_{j=1}^{\infty}x_j+1}
{\sum_{j=1}^{\infty}x_j+\theta}, \ \ \xb=\{x_i\}_{i\in\bn},\
\theta\in\bq_p, \ i\in\bn.
\end{equation}
\end{thm}

The proof consists of checking condition \eqref{comp} for the
measures \eqref{mu} (cp. \cite{GR}).

\begin{rem}\label{1} Note that thanks to the non-Archimedeanity of the
norm $|\cdot|_p$ the series $\sum\limits_{k=1}^\infty x_k$
converges iff the sequence $\{x_n\}$ converges to 0 (see
\cite{Ko,S}). Therefore, from \eqref{H} one can see that
condition \eqref{L} and $|\exp_p(x)|_p=1$ imply that the series
$\sum\limits_{j=1}^\infty \hat h_{j,x}$ always converges and is
finite.
\end{rem}

\section{Absence of the phase transition}

In this section, under some condition on weights $\l$, we are
going to prove the absence of the phase transition for the model
\eqref{Potts}

As pointed out from \eqref{H},\eqref{L} we have  $|\hat h_n|_p\to
0$ as $n\to\infty$. Therefore, let us consider the following space
$$
c_0=\{\xb=\{x_n\}_{n\in\bn}\in\bq_p^\bn: \ |x_n|_p\to 0, \
n\to\infty\}
$$
with a norm $\|\xb\|=\max\limits_n|x_n|_p$ (see \cite{Ro,S} for
more $p$-adic Banach spaces). Put $\Bb=\{\xb\in c_0: \
\|\xb\|<1\}$. Since the norm takes discrete values, the set $\Bb$
coincides with $\Bb=\{\xb\in c_0: \ \|\xb\|\leq 1/p\}$. Therefore,
$\Bb$ is a closed set.

Let us for the sake of shortness, given a sequence
$\xb=\{x_n\}_{n\in\bn}$, denote
\begin{equation}\label{x1}
X:=\sum_{j=1}^\infty x_j.
\end{equation}

Then one can easily see that
\begin{equation}\label{xy}
|X-Y|_p=\bigg|\sum_{j=1}^\infty
(x_j-y_j)\bigg|_p\leq\max_j|x_j-y_j|=\|\xb-\yb\|,
\end{equation}
 for any $\xb,\yb\in c_0$.

\begin{lem}\label{frac} For the mapping $F_i$ given by \eqref{eq2} the following
relations hold
\begin{eqnarray}\label{gxy}
&&|F_i(\xb,\t)-F_i(\yb,\t)|_p\leq|\t-1|_p\|\xb-\yb\|,\\
\label{gx} && |F_i(\xb,\t)-1|_p=|\t-1|_p.
\end{eqnarray}
for every $\xb,\yb\in \Bb$ and $i\in\bn$.
\end{lem}

\begin{pf} Let $\xb,\yb\in\Bb$, then  from \eqref{eq2} with \eqref{xy} we have
\begin{eqnarray*}
|F_i(\xb,\t)-F_i(\yb,\t)|_p&=&|((\t-1)x_i+X+1)(Y+\t)-((\t-1)y_i+Y+1)(X+\t)|_p\\
&=&|\t-1|_p|x_iY-y_iX+\t(x_i-y_i)+X-Y|_p\\
&=&|\t-1|_p\bigg|(X+\t)(x_i-y_i)+(1-x_i)(X-Y)|_p\\
&\leq&|\t-1|_p\max\{|x_i-y_i|_p,|X-Y|_p\}\\
&\leq&|\t-1|_p\|\xb-\yb\|
\end{eqnarray*}
which proves \eqref{gxy}, here we have used that
$|X+\t|_p=1$,$|Y+\t|_p=1$. Next relation \eqref{gx} is obvious.
\end{pf}

Let us first enumerate $S(x)$ for any $x\in V$ as follows
$S(x)=\{x_1,\cdots,x_k\}$, here as before $S(x)$ is the set of
direct successors of $x$ (see \eqref{S(x)}). Using this
enumeration one can rewrite \eqref{eq1} by
\begin{eqnarray}\label{eq11}
\hat h_{i,x}=\frac{\l(i)}{\l(0)}\prod_{m=1}^k
F_i(\hat\h_{x_m};\t_{x,x_m}), \ \ i\in\bn, \ \ \textrm{for every}
\ \ x\in V\setminus\{x^{(0)}\}.
\end{eqnarray}

Now we need an auxiliary fact.

\begin{lem}\label{pr} If $|a_i|_p\leq 1$, $|b_i|_p\leq 1$, $i=1,\dots,n$, then
\begin{equation*}
\bigg|\prod_{i=1}^{n}a_i-\prod_{i=1}^n b_i\bigg|_p\leq \max_{i\leq
i\leq n}\{|a_i-b_i|_p\}
\end{equation*}
\end{lem}

\begin{pf} We have
\begin{eqnarray*}
\bigg|\prod_{i=1}^{n}a_i-\prod_{i=1}^n b_i\bigg|_p&\leq&
\bigg|a_1\bigg(\prod_{i=2}^{n}a_i-\prod_{i=2}^n
b_i\bigg)+(a_1-b_1)\prod_{i=2}^nb_i\bigg|_p\\
&\leq&
\max\bigg\{|a_1-b_1|_p,\bigg|\prod_{i=2}^{n}a_i-\prod_{i=2}^n
b_i\bigg|_p\bigg\}\\
&\leq&\cdots\\
&\leq &\max_{i\leq i\leq n}\{|a_i-b_i|_p\}
\end{eqnarray*}
which is the assertion.
\end{pf}

So, we can formulate the main result.

\begin{thm}\label{uniq} Assume that a weight $\l$ satisfies the following condition
\begin{equation}\label{L1}
\max_i\bigg|\frac{\l(i)}{\l(0)}\bigg|_p<1.
\end{equation}
 Then there is no phase transition for the
countable state $p$-adic Potts model \eqref{Potts}, i.e.
$|\cs|\leq 1$.
\end{thm}

\begin{pf} From \eqref{L1} and \eqref{H} we see that all solutions
of \eqref{eq11} belong to $\Bb$. Now let us assume that
$\hat\h=\{\hat\h_x,x\in V\setminus\{x^{(0)}\}\},
\hat\sb=\{\hat\sb_x,x\in V\setminus\{x^{(0)}\}\}$ be the two
solutions of \eqref{eq11}. Now fix an arbitrary vertex $x\in
V\setminus\{x^{(0)}\}$. Then \eqref{eq11} with Lemmas \ref{frac}
and \ref{pr}, implies that
\begin{eqnarray*}
|\hat h_{i,x}-\hat
s_{i,x}|_p&=&\bigg|\frac{\l(i)}{\l(0)}\bigg|_p\bigg|\prod_{m=1}^k
F_i(\hat\h_{x_m};\t_{x,x_m})-\prod_{m=1}^k
F_i(\hat\sb_{x_m};\t_{x,x_m})\bigg|_p\\
& \leq & \bigg|\frac{\l(i)}{\l(0)}\bigg|_p\max_{1\leq m\leq
k}\bigg\{|F_i(\hat\h_{x_m};\t_{x,x_m}))_i-F_i(\hat\sb_{x_m};\t_{x,x_m})|_p\bigg\}\\
&\leq&\max_{1\leq m\leq
k}\bigg\{|\t_{x,x_m}-1|_p\|\hat\h_{x_m}-\hat\sb_{x_m}\|\bigg\}\\
&\leq &\frac{1}{p}\max_{1\leq m\leq
k}\bigg\{\|\hat\h_{x_m}-\hat\sb_{x_m}\|\bigg\}.
\end{eqnarray*}
Hence,
\begin{equation}\label{hx}
\|\hat\h_x-\hat\sb_x\|\leq \frac{1}{p}\max_{1\leq m\leq
k}\bigg\{\|\hat\h_{x_m}-\hat\sb_{x_m}\|\bigg\}
\end{equation}

Now take an arbitrary $\e>0$ and $n_0\in\bn$ such that
$1/p^{n_0}<\e$. Iterating \eqref{hx} $n_0$ times one gets
$\|\hat\h_x-\hat\sb_x\|<\e$. Therefore, the arbitrariness of $\e$
and $x$  yield that $\hat\h_x=\hat\sb_x$ for every $x\in
V\setminus\{x^{(0)}\}$. This means that $|\cs|\leq 1$ and
completes the proof.
\end{pf}

\begin{rem}  It is clear that the condition \eqref{L1} does not depend
on values of the prime $p$, therefore an analogues fact is not
true when the number of spins is finite ( see also Remark \ref{rb}
(a)).
\end{rem}

\begin{rem}  Note that the equality can be interpreted as an infinite dimensional
recurrence equation over the tree. So, Theorem \ref{uniq} means
that the equation has no more one solution. Simpler, recurrence
equations  over $p$-adic numbers were considered in
\cite{EPSW},\cite{M}.
\end{rem}

\section{Uniqueness and boundedness of the Gibbs measure}

As we pointed out, in general, $p$-adic Gibbs measures may not
exist. In this section we will show that the $p$-adic Gibbs
measure is unique under the condition \eqref{L1} for the
homogeneous model \eqref{Potts}.

Throughout this section we suppose that $J_{x,y}=J$. Recall that a
function $\h=\{\h_x, x\in V\setminus\{x^0\}\}$ is {\it
translation-invariant} if $\h_x=\h_y$ for every $x,y\in
V\setminus\{x^0\}$. We are going to show that \eqref{eq1} has a
translation invariant solution. To this end, consider the
following mapping $\cf:c_0\to\bq_p^\bn$ defined by
\begin{equation}\label{F}
(\cf(\xb))_i=\frac{\l(i)}{\l(0)}(F_i(\xb;\t))^k, \ \  \ i\in\bn,
\end{equation}
where $\xb=\{x_n\}\in c_0$. One can see that the domain of the
mapping is not whole space $c_0$\footnote{More exactly, the domain
 is $\{\xb\in c_0: \sum_{n=1}^\infty x_n+\t\neq 0\}$}, therefore,
in general, it is unbounded. But we are are going to examine $\cf$
on $\Bb$. Basically, for the mapping we do not have the inclusion
$\cf(\Bb)\subset\Bb$. However, from Lemma \ref{frac} and
\eqref{L1} we derive the following

\begin{lem}\label{inv} Let for a weight $\l$ condition \eqref{L1} be satisfied.
Then $\cf(\Bb)\subset \Bb$. Moreover, \begin{equation}\label{fxy}
\|\cf(\xb)-\cf(\yb)\|\leq|\t-1|_p\|\xb-\yb\| \ \ \textrm{for
every} \ \ \xb,\yb\in\Bb.
\end{equation}
\end{lem}

Now noting that $|\t-1|_p< 1/p^{1/(p-1)}$  and according to the
above Lemma, we can apply the fixed point theorem to $\cf$, which
implies the existence of a unique fixed point $\hat\h_\l=\{\hat
h_{\l,n}\}_{n\in\bn}\in\Bb$ (here the solution depends on a weight
$\l$, therefore, we indicate that dependence by subscript $\l$).
Since $\hat \h_{\l}$ is a fixed point of $\cf$ one has
\begin{equation}\label{x0h}
\frac{\l(0)}{\l(i)}\hat h_{\l,i}=(F_i(\hat\h_\l;\t))^k,
\end{equation}
which, thanks to \eqref{gx} and Lemma \ref{pr}, implies that
\begin{eqnarray*}
&&\bigg|\frac{\l(0)}{\l(i)}\hat h_{\l,i}\bigg|_p=1, \ \ \ \
\bigg|\frac{\l(0)}{\l(i)}\hat h_{\l,i}-1\bigg|_p\leq|\t-1|_p
\end{eqnarray*}
Therefore, Lemma \ref{exp} allows us to take logarithm from both
sides of \eqref{H}, and we obtain
\begin{equation*}
h_{\l,i}-h_{\l,0}=\log_p\bigg(\frac{\l(0)}{\l(i)}\hat
h_{\l,i}\bigg),
\end{equation*}
which, due to Theorem \ref{comp1}, defines the $p$-adic Gibbs
measure, which is denoted by $\m_\l$. Now combining this with
Theorem \ref{uniq} we have the following

\begin{thm}\label{uniq1} Let $0<|J|_p<1/p^{1/(p-1)}$ and for a weight $\l$ condition \eqref{L1} be satisfied.
Then for homogeneous  $p$-adic  Potts model \eqref{Potts} on the
Cayley tree of order $k$ there is a unique $p$-adic Gibbs measure
$\mu_\l$.
\end{thm}

\begin{rem}\label{rb} It is  worth to emphasize the following notes:

\begin{itemize}

\item[(a)]  Note that  in \cite{MR2} we have proven for the
$q$-state Potts model that the $p$-adic Gibbs measure is unique if
$q$ and $p$ are relatively prime. Therefore, the proven Theorem
\ref{uniq} shows the difference between finite and countable state
Potts models.

\item[(b)] It turns out that condition \eqref{L1} is important. If
we replace strict inequality there with weaker one $\leq$, then
Theorem \ref{uniq} may not hold. Namely, in that case there may
occur a phase transition.  Indeed, if $\l(0)=\l(1)=\l(2)=1$ and
$\l(k)=0$ for every $k\geq 3$, then clearly  \eqref{L1} is not
satisfied. On the other hand, our model is reduced to 3-state
Potts model. For such a model in \cite{MR1} the existence of the
phase transition was proven at $p=3$. Moreover, in that case,
$p$-adic Gibbs measures were unbounded.

 \item[(c)] For the real counterpart of the model
uniqueness result is still unknown (see \cite{Ga,GR}).

\end{itemize}
\end{rem}

Now we propose the following

{\bf Problem.} Investigate all fixed points and behavior around
such points of the dynamical system \eqref{F} on whole space
$c_0$. Note that the dynamical system is rational. Therefore, we
hope that general theory of rational dynamical systems developed
in \cite{B1,B2,B3,Be,RL,Sil2}, can be applied for one.\\

To establish boundedness of the measure $\m_\l$, we need the
following auxiliary result.

\begin{lem} Let $\h$ be a solution of \eqref{eq1}, and
$\m_\h$ be an associated $p$-adic Gibbs measure. Then for the
corresponding partition function $Z^{(\h)}_n$ (see \eqref{ZN1})
the following equality holds
\begin{equation}\label{ZN2}
Z^{(\h)}_{n+1}=A_{\h,n}Z^{(\h)}_n,
\end{equation}
where $A_{\h,n}$ will be defined below (see \eqref{aN3}).
\end{lem}

\begin{pf} Since $\h$ is a solution of \eqref{eq1}, then we conclude that there is a constant
$a_\h(x)\in\bq_p$ such that
\begin{equation}\label{aN1}
\prod_{y\in
S(x)}\sum_{j\in\Phi}\exp_p\{J\d_{ij}+h_{j,y}\}\l(j)=a_{\h}(x)\exp_p\{h_{i,x}\}
\end{equation}
for any $i\in\Phi$. From this one gets
\begin{eqnarray}\label{aN2}
\prod_{x\in W_{n}}\prod_{y\in
S(x)}\sum_{j\in\Phi}\exp_p\{J\d_{ij}+h_{j,y}\}\l(j)&=&\prod_{x\in
W_n}a_{\h}(x)\exp_p\{h_{i,x}\}\nonumber\\
&=&A_{\h,n}\exp_p\bigg\{\sum_{x\in W_n}h_{i,x}\bigg\},
\end{eqnarray}
where
\begin{equation}\label{aN3}
A_{\h,n}=\prod_{x\in W_n}a_{\h}(x).
\end{equation}
By  \eqref{mu},\eqref{aN2} we have
\begin{eqnarray*}
1&=&\sum_{\s\in\Om_n}\sum_{\w\in\Phi}\m^{(n+1)}_\h(\s\vee\w)\\
&=&\sum_{\s\in\Om_n}\sum_{\w\in\Phi}\frac{1}{Z^{(\h)}_{n+1}}\exp_p\bigg\{H(\s\vee\w)+\sum_{x\in
W_{n+1}}h_{\w(x),x}\bigg\}\prod_{x\in V_{n}}\l(\s(x))\prod_{y\in W_{n+1}}\l(\w(y))\\
&=&\frac{1}{Z^{(\h)}_{n+1}}\sum_{\s\in\Om_n}\exp_p\{H(\s)\}\prod_{x\in
V_n}\l(\s(x))\prod_{x\in W_n}\prod_{y\in S(x)}\sum_{j\in\Phi}\exp_p\{J\d_{\s(x),j}+h_{j,y}\}\l(j)\\
&=&\frac{A_{\h,n}}{Z^{(\h)}_{n+1}}
\sum_{\s\in\Om_n}\exp_p\bigg\{H(\s)+\sum_{x\in W_n}h_{\s(x),x}\bigg\}\prod_{x\in V_n}\l(\s(x))\\
&=&\frac{A_{\h,n}}{Z^{(\h)}_{n+1}}Z_n^{(\h)}
\end{eqnarray*}
which implies the required relation.
\end{pf}

Now, as before, assume that condition \eqref{L1} is satisfied.
Then we know that the equation \eqref{eq1} has a unique
translation invariant solution $\hat\h_\l=\{\hat h_{\l,n}\}$,
therefore, $a_\h(x)$ does not depend on $x$, which will be denoted
by $a$. Hence, from \eqref{aN1} one finds
\begin{equation}\label{aN4}
a=\exp_p\{(k-1)h_{\l,0}\}(\l(0))^k\bigg(\t+\sum_{j=1}^\infty \hat
h_{\l,j}\bigg)^k.
\end{equation}
The equalities \eqref{ZN2} and \eqref{aN3} imply that
\begin{equation}\label{ZN3}
Z_{\l,n}=a^{|V_{n-1}|},
\end{equation}
where $Z_{\l,n}$ denotes the partition function of the measure
$\mu_\l$ corresponding to the unique solution.
 Now we are ready to formulate a result.

\begin{thm}\label{bound} Assume that \eqref{L1} is satisfied. Then
the $p$-adic Gibbs measure $\m_\l$ is bounded.
\end{thm}

\begin{pf} Take any $\s\in\Om_{V_n}$. Then from
\eqref{mu} with \eqref{ZN3}, \eqref{L1} one gets
\begin{eqnarray}\label{m0b}
|\m_\l(\s)|_p&=&\frac{1}{|Z_{\l,n}|_p}\bigg|\exp_p\{H(\s)+\sum_{x\in W_n}h_{\l,\s(x)}\}\prod_{x\in V_n}\l(\s(x))\bigg|_p\nonumber\\
&=&\frac{1}{|\l(0)|_p^{k|V_{n-1}|}|\t+\sum_{j=1}^\infty\hat h_{\l,j}|_p^k}\prod_{x\in V_n}|\l(\s(x))|_p\nonumber\\
&=&\frac{|\l(0)|_p^{|V_n|}}{|\l(0)|_p^{k|V_{n-1}|}}\prod_{x\in V_n}\bigg|\frac{\l(\s(x))}{\l(0)}\bigg|_p\nonumber\\
&\leq &|\l(0)|_p^{|V_n|-k|V_{n-1}|},
\end{eqnarray}
here we have used that $|\t+\sum_{j=1}^\infty\hat h_{\l,j}|_p=1$.
It is known \cite{B} that
$$
|V_n|=1+\frac{k+1}{k-1}(k^n-1),
$$
therefore,
\begin{equation*}
|V_n|-k|V_{n-1}|=2
\end{equation*}
hence, \eqref{m0b} implies that $\m_\l$ is bounded.
\end{pf}

\begin{ex} Assume that a solution of \eqref{eq1} is
$\hat h_i=p^i$, $i\in\bn$. In this case, one can see that
\begin{equation*}
\sum_{j=1}^\infty \hat h_j=\frac{p}{1-p}.
\end{equation*}
Let us find the corresponding weight $\l$. Put $\l(0)=1$, then
from \eqref{eq1} one gets
\begin{equation*}
\l(n)=p^n\bigg(\frac{p(1-\t)+\t}{(\t-1)p^n(1-p)+1}\bigg)^k, \ \
n\in\bn
\end{equation*}
which evidently satisfies conditions \eqref{L} and \eqref{L1}. So,
there is a unique bounded $p$-adic Gibbs measure on $\Om$.
\end{ex}

\section{Certain properties of the $p$-adic Gibbs measures}

In this section without loss of generality we assume that for
weights $\l(0)=1$ and $h_0=0$.

Denote
\begin{equation*}
\cw=\{\{\l(i)\}_{i\in\Phi}\in\bq_p^\Phi: \ \l(0)=1,\ |\l(i)|_p<1,\
|\l(n)|_p\to 0\ \textrm{as} \ n\to\infty\}.
\end{equation*}
A norm of a weight $\l\in \cw$ we define by
$\|\l\|_\cw=\max\limits_{n\in\Phi}\{|\l(n)|_p\}$, therefore, it is
clear that $\|\l\|_\cw=1$. We know from Theorems \ref{uniq1} and
\ref{bound} that for every $\l\in\cw$ there is a unique bounded
$p$-adic Gibbs measure $\m_\l$, by denote $\cg_P$ the set of such
measures corresponding to the homogeneous Potts model
\eqref{Potts}. We endow $\cg_P$ with a norm defined by
\begin{equation}\label{Nm}
\|\m\|_\cg=\max_{\s\in\Om_{V_n}\atop n\in\bn}|\m(\s)|_p, \ \
\m\in\cg_P.
\end{equation}

One can ask: does the measure $\m_\l$ depend on $\l$ continuously?
Next Theorem gives an answer to the question.

\begin{thm}\label{mmap} For every $\l,\k\in\cw$ one has
\begin{equation}\label{mm1}
\|\m_\l-\m_\k\|_\cg\leq\|\l-\k\|_\cw.
\end{equation}
Hence, the correspondence $\l\mapsto\m_\l$ is continuous.
\end{thm}

\begin{pf} Take any $\l,\k\in\cw$. Let $\hat\h_\l=\{\hat h_{\l,i}\}_{i\in\bn}$ and
$\hat\h_\k=\{\hat h_{\k,i}\}_{i\in\bn}$ be the corresponding
solutions of \eqref{eq1}\footnote{Here we again recall that
according to Theorems \ref{uniq} and \ref{uniq1} such solutions
exist.}. Denote $\h_{\l}=\{h_{\l,i}=\log_p\hat h _{\l,i}\},$
$\h_{\k}=\{h_{\k,i}=\log_p\hat h _{\k,i}\}$, which exist due to
the proof of Theorem \ref{uniq1}. Using \eqref{gxy} consider the
difference
\begin{eqnarray}\label{xlk1}
|\hat h_{\l,i}-\hat h_{\k,i}|_p&=&|\l(i)F_i(\hat \h_{\l};\t)-\k(i)F_i(\hat \h_{\k};\t)|_p\nonumber \\
&=&|\l(i)(F_i(\hat \h_{\l};\t)-F_i(\hat \h_{\k};\t))+(\l(i)-\k(i))F_i(\hat \h_{\k};\t)|_p\nonumber\\
&\leq&\max\bigg\{|\l(i)|_p|F_i(\hat \h_{\l};\t)-F_i(\hat \h_{\k};\t)|,|\l(i)-\k(i)|\bigg\}\nonumber\\
&\leq&\max\bigg\{|\l(i)|_p|\t-1|_p\|\hat \h_{\l}-\hat
\h_{\k}\|,|\l(i)-\k(i)|\bigg\}.
\end{eqnarray}
Taking into account $|\l(i)|_p|\t-1|_p\|\hat \h_{\l}-\hat
\h_{\k}\|<\|\hat \h_{\l}-\hat \h_{\k}\|$ from \eqref{xlk1} and
non-Archimedeanity of the norm $\|\cdot\|$ one has
\begin{equation}\label{xlk2}
\|\hat \h_{\l}-\hat \h_{\k}\|=\|\l-\k\|_\cw.
\end{equation}

Now take any $n\in\bn$. Let us estimate the difference between the
partition functions $Z_{\l,n}$ and $Z_{n,\k}$ (see \eqref{ZN1}) of
the measures $\m_\l$ and $\m_\k$, respectively. Thanks to
\eqref{ZN3} with \eqref{aN4} and \eqref{xy} one gets
\begin{eqnarray}\label{Zlk}
|Z_{\l,n}-Z_{\k,n}|_p&=&\bigg|\bigg(\t+\sum_{j=1}^\infty \hat
h_{\l,j}\bigg)^k-\bigg(\t+\sum_{j=1}^\infty
\hat h_{\k,j}\bigg)^k\bigg|_p\nonumber\\
&\leq&\bigg|\sum_{j=1}^\infty
\hat h_{\l,j}-\sum_{j=1}^\infty \hat h_{\l,j}\bigg|_p\nonumber\\
&\leq &\|\hat \h_\l-\hat \h_\k\|
\end{eqnarray}

Hence, for any $\s\in\Om_{V_n}$,  from \eqref{mu} and Lemma
\ref{pr} using \eqref{xlk2},\eqref{Zlk} we obtain
\begin{eqnarray}\label{mlk1}
|\m_\l(\s)-\m_\k(\s)|_p&=&\bigg|\frac{1}{Z_{\l,n}}\exp_p\bigg\{\sum_{x\in
W_n} h_{\l,\s(x)}\bigg\}\prod_{u\in V_n}\l(\s(u))\nonumber \\
&&- \frac{1}{Z_{\k,n}}\exp_p\bigg\{\sum_{x\in W_n}
h_{\k,\s(x)}\bigg\}\prod_{u\in
V_n}\k(\s(u))\bigg|_p\nonumber\\
&=&\bigg|\frac{1}{Z_{\l,n}}\prod_{x\in W_n}
\hat h_{\l,\s(x)}\prod_{u\in V_{n-1}}\l(\s(u))\nonumber\\
&&-\frac{1}{Z_{\k,n}}\prod_{x\in W_n} \hat h_{\k,\s(x)}\prod_{u\in
V_{n-1}}\k(\s(u))\bigg|_p\nonumber\\
&\leq
&\max_{x\in W_n\atop u\in V_{n-1}}\bigg\{\bigg|\frac{1}{Z_{\l,n}}-\frac{1}{Z_{\k,n}}\bigg|_p,|\hat h_{\l,\s(x)}-\hat h_{\k,\s(x)}|_p,\nonumber\\
&&|\l(\s(u))-\k(\s(u))|_p\bigg\}\nonumber\\
&\leq
&\max\bigg\{|Z_{\l,n}-Z_{\k,n}|_p,\|\hat \h_{\l}-\hat \h_{\k}\|,\|\l-\k\|_\cw\bigg\}\nonumber\\
&\leq&\|\l-\k\|_\cw,\end{eqnarray} here we have used  equalities
$|Z_{\l,n}|_p=1$, $|Z_{\k,n}|_p=1$, which come from \eqref{ZN3}.

Due to the arbitrariness of $n$ and $\s$ we get the required
relation \eqref{mm1}. \end{pf}

Now consider one limit theorem concerning $\m_\l$.

Let us fix a wight $\l\in \cw$ such that $\l(i)\neq 0$ for all
$i\in\bn$. As before by $\hat \h_\l$ we denote a solution of
\eqref{eq1}, and, as before, $h_{\l,i}=\log_p \hat h_{\l,i}$,
$i\in\bn$.

Let us denote
\begin{equation}\label{eA}
A_n=\bigg\{\s\in\Om_{V_n}: J\sum_{<x,y>\in
V_n}\d_{\s(x),\s(y)}+\sum_{x\in W_n}h_{\l,\s(x)} \equiv 0(mod\
p^n),\ \bigg\}.
\end{equation}

By $ \l^{\otimes,n}$ denote the following measure defined on
$\Om_{V_n}$
\begin{equation}\label{lmu}
\l^{\otimes, n}(\s)=\frac{1}{Z_{\l,n}}\prod_{x\in V_n}\l(\s(x)), \
\ \s\in\Om_{V_n}.
\end{equation}

\begin{thm}\label{limit} Let $\m_\l$ be the $p$-adic Gibbs measure
corresponding to the Potts model with a weight $\l$. Then one has
\begin{equation}\label{eA1}
\max_{\s\in
A_n}\bigg|\frac{\m_\l(\s)}{\l^{\otimes,n}(\s)}-1\bigg|_p\to 0 \ \
\textrm{as} \ \ n\to\infty.
\end{equation}
\end{thm}

\begin{pf} Note that if $\s\in A_n$ then it means that
\begin{equation*}
J\sum_{<x,y>\in V_n}\d_{\s(x),\s(y)}+\sum_{x\in
W_n}h_{\l,\s(x)}=Mp^n
\end{equation*}
for some $M\in\bz$. Therefore, the last equality  with
\eqref{lmu}, \eqref{mu} implies that
\begin{eqnarray*}
\bigg|\frac{\m_\l(\s)}{\l^{\otimes,n}(\s)}-1\bigg|_p&\leq&
\bigg|\exp_p\{Mp^n\}-1\bigg|_p\\
&=&|Mp^n|_p\\
&\leq& p^{-n}\to 0 \ \ \textrm{as} \ \ n\to\infty
\end{eqnarray*}
which proves the assertion.
\end{pf}

\begin{rem} In the theory of Markov process it is important to know whether a given
Markov measure or Markov field has some clustering (i.e. mixing)
property. It is known \cite{G} that, in the real case, when a
Gibbs measure is unique, then it basically has that property. In a
$p$-adic case, the situation is rather tricky (see \cite{Kh3,K4})
and still lots are not known. The proven Theorem shows that we are
able to find a set $A_n$ on which the $p$-adic measure $\m_\l$
becomes product measure when $n$ is large enough, which means
$\m_\l$ has a clustering property on $A_n$. The results in
\cite{Kh1,Kh3} show that, in general, the clustering property does
not hold for $p$-adic measures.
\end{rem}

\begin{rem} Note that in \cite{K4} the limit
behaviour of sums of independent equally distributed random
variables with respect to $p$-adic Bernoulli measure has been
studied. The measure $\m_\l$ defines a Markov process with
infinite number of states, so the last theorem is the first step
in an investigation of limit theorems for dependent processes in a
$p$-adic context.
\end{rem}

\section{conclusions}

Investigations of physical models over $p$-adic field require a
new kind of probability theory \cite{K3},\cite{KhN}. Development
of the $p$-adic probability theory gives a possibility to study
$p$-adic statistical mechanics models. In the present paper we
have studied countable state nearest-neighbor $p$-adic Potts
models on a Cayley tree in the $p$-adic probability scheme. For
the the model, we gave a construction of $p$-adic Gibbs measures
which depends on weight $\l$. Such measures are natural and
provide nontrivial concrete examples of $p$-adic Markov processes
with countable state space (see \cite{KL}). We have shown that
under some condition on weights for the model, the absence of a
phase transition by studying an infinite-dimensional recursion
equation. Such results are unknown for the real counterparts of
the considered models (see \cite{Ga,GR}). It turned out that the
condition does not depend on values of the prime $p$, therefore an
analogous fact is not true when the number of spins is finite
\cite{MR1,MR2}. It has been proven that $p$-adic Gibbs measure for
homogeneous Potts model is unique. We also established boundedness
such a measure. Continuous dependence the measure on weights was
also proven. We even obtain one limit theorem for such a measure.
This is the first step in an investigation on limit theorems for
dependent processes in a $p$-adic context.

\section*{Acknowledgement} The author F.M. thanks the FCT (Portugal)
grant SFRH/BPD/17419/2004. He is also grateful to Prof. A.Yu.
Khrennikov at  V\"{a}xj\"{o} University for kind hospitality. The
third named author J.F.F.M. thanks DYSONET-project for partial
support. Finally, the authors also would like to thank to the
referee for his useful suggestions which allowed us to improve the
text of the paper.


\end{document}